\documentclass[showpacs,amsmath,amssymb]{revtex4}

\usepackage{graphicx}
\usepackage{dcolumn}
\usepackage{bm}

\begin{document}

\title{
{\bf{Importance of $1n$-stripping process in the $^{6}$Li+$^{159}$Tb reaction}}
}

\author{M. K. Pradhan$^{1}$}
\author{A. Mukherjee$^{1}$}
\email{anjali.mukherjee@saha.ac.in}
\author{Subinit Roy$^{1}$}
\author{P. Basu$^{1}$}
\author{A. Goswami$^{1}$} 
\author{R. Kshetri$^{1}$}
\author{R. Palit$^{2}$}
\author{V. V. Parkar$^{3}$}
\author{M. Ray$^{4}$} 
\author{M. Saha Sarkar$^{1}$} 
\author{S. Santra$^{3}$}

\affiliation{${^1}$ Nuclear Physics Division, Saha Institute of Nuclear 
Physics, 1/AF, Bidhan Nagar, Kolkata-700064, India }
\affiliation{${^2}$ Department of Nuclear $\&$ Atomic Physics, Tata Institute 
of Fundamental Research, Mumbai-400005, India }
\affiliation{${^3}$ Nuclear Physics Division, Bhabha Atomic Research Centre, 
Mumbai-400085, India }
\affiliation{${^4}$ Department of Physics, Behala College, Parnasree, 
Kolkata-700060, India}

\begin{abstract}
       
        The inclusive cross sections of the $\alpha$-particles produced in the
reaction $^{6}$Li+$^{159}$Tb have been measured at energies around the Coulomb 
barrier. The measured cross sections are found to be orders of magnitude larger
than the calculated cross sections of $^{6}$Li breaking into $\alpha$ and $d$
fragments, thus indicating contributions from other processes. The experimental 
cross sections of $1n$-stripping and $1n$-pickup processes have been determined
from an entirely different measurement, reported earlier. Apart from incomplete
fusion and/ $d$-transfer processes, the $1n$-stripping process is found to be a
significant contributor to the inclusive $\alpha$-particle cross sections in 
this reaction. 
\end{abstract}

\pacs{24.10.Eq, 25.70.Jj, 25.60.Pj, 25.70.Mn, 27.70.+q}


\maketitle

\section{Introduction}

   Investigation of reactions involving weakly bound projectiles and  
the influence of their low binding energies on various reaction channels has 
received a fillip in recent years, especially in the context of the 
increasing number of radioactive ion beam facilities. To have a proper 
understanding of the influence of breakup of loosely bound projectiles on the 
fusion process, one needs to understand the mechanisms of all the competing 
reaction channels.

    Measurements involving weakly bound projectiles, both stable and unstable, 
with $\alpha$+x cluster structures show substantially large production cross 
sections for $\alpha$-particles [1-9], which indicate the presence of 
mechanisms other than the $\alpha$+x breakup. Utsunomiya {\em{et al.}} showed 
that for the reaction $^{7}$Li+$^{159}$Tb \cite{Utsu83}, about half of the 
$\alpha$ and triton yield originates from the breakup-fusion process, which is
more commonly referred to as the incomplete fusion (ICF) process. Evidence of 
transfer-induced breakup producing $\alpha$-particles in the reaction 
$^{7}$Li+$^{65}$Cu has also been reported \cite{Shriv06}. Our recent works on 
the systematic measurements of complete and incomplete fusion excitation 
functions for the reactions $^{6,7}$Li+$^{159}$Tb and $^{10,11}$B+$^{159}$Tb 
\cite{Mukh06,Mukh10,Pra11} have shown that the complete fusion (CF) cross 
sections at above-barrier energies are suppressed for reactions with weakly 
bound projectiles, and the extent of suppression is correlated with the 
$\alpha$-breakup threshold of the projectile. The measurements also showed 
that the $\alpha$-emitting channel is the favoured ICF process in reactions 
with projectiles having low $\alpha$-breakup thresholds.  A critical insight 
into these measurements shows that the sum of the CF and the ICF cross sections
for each system yields the total fusion cross sections which lie very close to 
the calculated one dimensional Barrier Penetration Model calculations, at 
energies above the barrier. This shows that the suppression in the CF cross 
sections at above-barrier energies is primarily due to the loss of flux into 
the ICF channel.
     
         A recent exclusive  measurement on the reaction $^{6}$Li+$^{208}$Pb 
\cite{Luong11} showed that the cross sections of the breakup process following 
$1n$-stripping (transfer-breakup) of $^{6}$Li are higher than that for the 
breakup of $^{6}$Li into $\alpha$ and {\em{d}} fragments. By contrast, another
recent work on the reaction $^{6}$Li+$^{209}$Bi \cite{Santra12}, aimed at 
disentangling the reaction mechanisms responsible for the large inclusive 
$\alpha$-particle cross sections, indicated that the cross sections of the 
breakup of $^{6}$Li into $\alpha$ and {\em{d}} fragments are much higher than 
those of the breakup following $1n$- stripping of $^{6}$Li. However, very 
recently it has been reported \cite{Luong13} that for $^{6,7}$Li induced 
reactions with $^{207,208}$Pb and $^{209}$Bi targets, projectile breakup is 
triggered predominantly by nucleon transfer, $n$-stripping for $^{6}$Li and 
$p$-pickup for $^{7}$Li. Based on the observations made in a few reactions, it 
will perhaps be too optimistic to generalize the dominance of transfer induced
breakup for all $^{6,7}$Li induced reactions, as the importance of a 
transfer reaction depends largely on the projectile-target combination. To 
conclude whether the observation is a general feature of $^{6,7}$Li induced 
reactions or is true only for specific reactions, it is important to carry 
out a systematic investigation of $^{6,7}$Li induced reactions on various 
targets, especially medium and light mass targets. In the background of this 
scenario we chose to carry out an inclusive measurement of the 
$\alpha$-particles produced at energies around the Coulomb barrier in the 
$^{6}$Li induced reaction with a $^{159}$Tb target. The reaction was so chosen 
because detailed CF and ICF cross sections have already been measured for the 
system \cite{Pra11}.  

\section{Experimental Details}
The $^{6}$Li beam with energies $E_{lab}$=23, 25, 27, 30 and 35 MeV, from the 
14UD BARC-TIFR Pelletron Accelerator Centre in Mumbai, was used to impinge a  
self-supporting $^{159}$Tb target foil of thickness $\sim$450 $\mu$g/cm$^{2}$.
The beam energies were corrected for loss of energy in the target material at 
half-thickness of the target. 
To detect and identify the $\alpha$-particles produced in the reaction, four 
$\Delta$E-E telescopes of Si-surface barrier detectors were placed on a movable
arm inside a scattering chamber of 1 m diameter. The thicknesses of the 
detectors were so chosen that the $\alpha$-particles lose part of their kinetic
energies in the first detector ($\Delta$E) and are stopped in the second 
detector (E$_{res.}$). The $\alpha$-particles produced in the reaction were 
measured in the range 30$^{\circ}$ $\leq {\theta}_{lab} \leq$165$^{\circ}$ in 
steps of 2$^{\circ}$ or 5$^{\circ}$ depending on the bombarding energy, where
${\theta}_{lab}$ is the scattering angle in laboratory. Two Si-surface 
barrier detectors, each of thickness 500 $\mu$m, were placed at angles of $\pm$
20$^{\circ}$ with-respect-to the beam direction for beam monitoring and 
normalization purposes. 

Figure 1(a) shows a typical two-dimensional inclusive $\Delta$E$-$E 
(E=$\Delta$E+E$_{res.}$) spectrum taken at the laboratory scattering angle,
${\theta}_{lab}$=99.5$^{\circ}$ for a beam energy of 27 MeV. The enclosed area 
in the figure shows the $\alpha$-particle band and its one-dimensional 
projection is shown in Fig.1(b). It shows a broad continuous peak, with 
centroid nearly equal to 2/3 times the incident beam energy. The contribution 
of the $\alpha$-particles, emitted mostly at energies corresponding to the beam
velocity, is expected to originate from breakup related processes. It needs to 
be mentioned here that the heavy compound nuclei formed, following either the 
CF or ICF process, are expected to decay predominantly by neutron evaporation 
\cite{Pra11} and this is also predicted by the statistical model calculations 
done using the code PACE2 \cite{Ga80}. 
\begin{figure}
\vspace{-50pt}
\begin{center}
\includegraphics[scale=0.7]{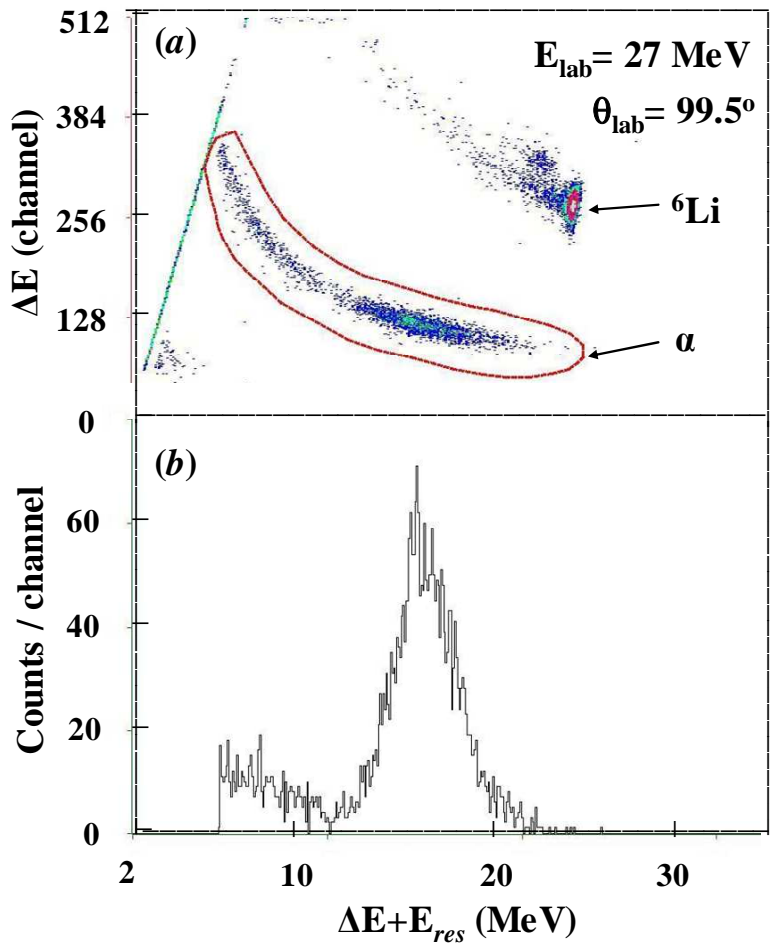}
\end{center}
\vspace{-140pt}
\caption{\label{fig:Fig1} (Color online) a) Typical  
two-dimensional $\Delta$E-E spectrum of the reaction $^{6}$Li+$^{159}$Tb at
scattering angle, $\theta_{lab}$=99.5$^{\circ}$ for the beam energy 27 MeV. 
The enclosed area shows the $\alpha$-particle band. b) The one-dimensional 
projection of the enclosed area in the upper figure (a).}
\end{figure}
      The differential cross sections of the inclusive $\alpha$-particles were
obtained by using the formula,
\begin{equation}
        \frac{d\sigma}{d\Omega}=\left (\frac{Y_{\alpha}}{{Y_{mon}}}\right )
\left (\frac{\Delta\Omega_{mon}}{\Delta\Omega_{Tele}}\right )
\left (\frac{d\sigma}{d\Omega}\right )^{Ruth}_{\theta_{mon}}
\end{equation}                             
where $Y_{\alpha}$ and $Y_{mon}$ are the number of counts under the broad 
continuous peak of the $\alpha$-particles (Fig.1) and the average number of 
counts in the monitor detectors, respectively. The quantities 
$\Delta\Omega_{mon}$ and $\Delta\Omega_{Tele}$ are the solid angles subtended 
by the monitor detectors and the $\Delta$E-E telescope, respectively and 
$\theta_{mon}$ is the angle of the monitor detector. For all the five 
bombarding energies, the broad peak in each of the $\alpha$-particle energy 
spectra was well separated from the low-energy small peak (Fig.1), at all 
scattering angles. The $\alpha$-particles in the low-energy peak, which is 
indeed a very small contribution at all the bombarding energies, could be due 
to target impurities, such as C and O.  

        The measured angular distributions of the inclusive $\alpha$-particles 
for the five incident energies are shown in Fig.2. The angular distribution at 
each of the bombarding energies was obtained by considering the counts within 
the main peak of the $\alpha$-spectrum. With the exception of the low energy 23
MeV data, each of the distributions shows a clear maximum that shifts to lower 
laboratory angle with the increase of beam energy. The angular distribution 
data were fitted with Gaussian functions and are shown by the lines in Fig.2. 
The total angle-integrated $\alpha$-particle cross sections obtained from the 
angular distribution data at each of the incident energies are plotted in Fig.3. 
\begin{figure}
\centering
\vspace{-170 pt}
\hspace{-0.8 in}
\begin{center}
\includegraphics[scale=0.79]{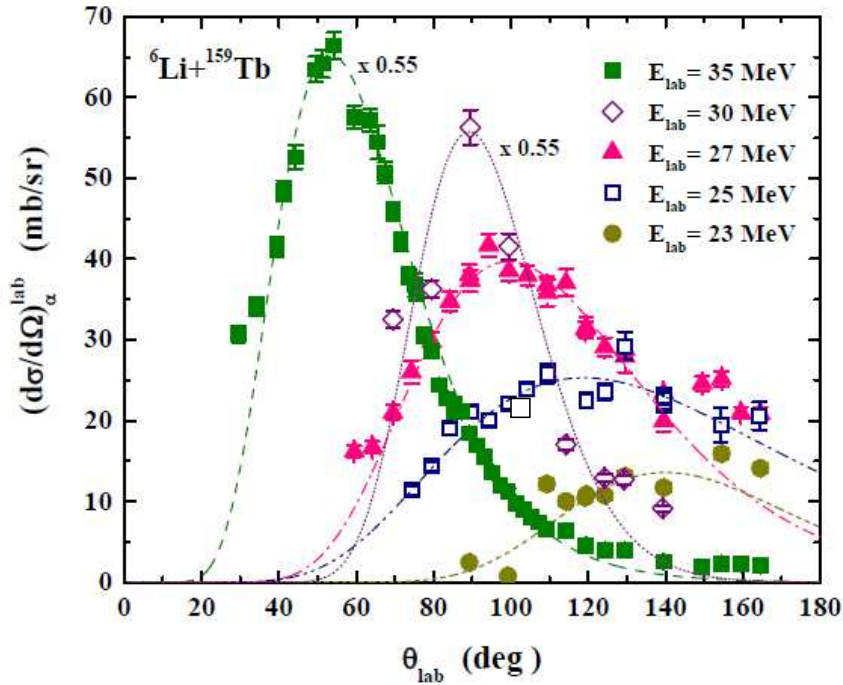}
\end{center}
\vspace{-180pt}
\caption{\label{fig:Fig2}(Color online)  
Angular distributions of inclusive $\alpha$-particles for the reaction 
$^{6}$Li+$^{159}$Tb at energies $E_{lab}$=23-35 MeV. The lines through the data
are fits with Gaussian functions.} 
\end{figure}
\begin{figure}
\centering
\vspace{-70 pt}
\hspace{-0.7 in}
\begin{center}
\includegraphics[scale=0.7]{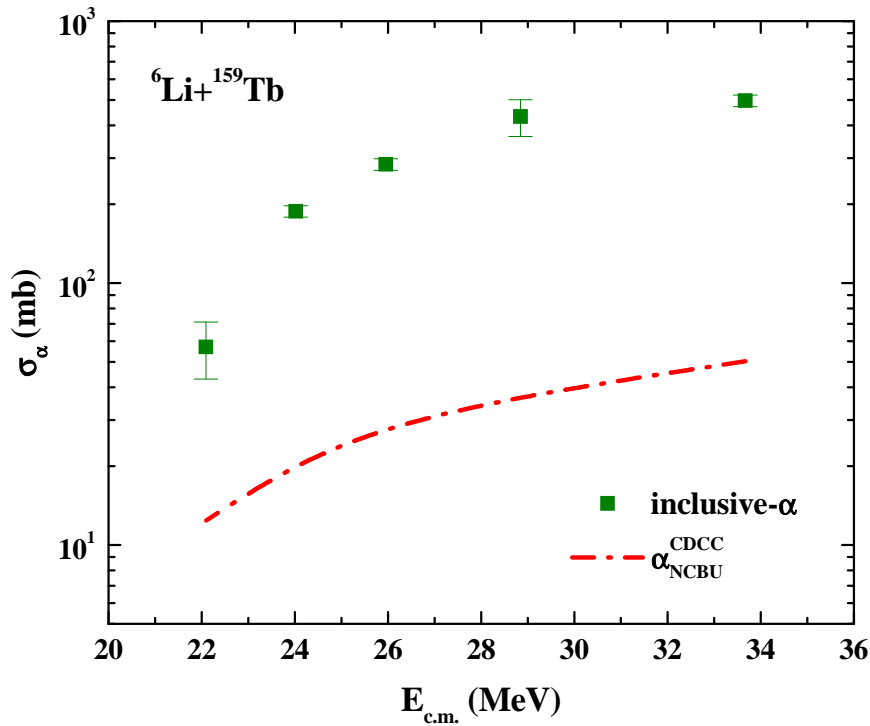}
\end{center}
\vspace{-190pt}
\caption{\label{fig:Fig3}(Color online)  
Measured inclusive $\alpha$-particle cross sections for the reaction 
$^{6}$Li+$^{159}$Tb. The dash-dotted line shows the NCBU cross sections 
obtained from the CDCC calculations.}
\end{figure}
\section{Discussion}
   Because the present work is an inclusive measurement, the $\alpha$-particle 
cross sections are expected to have contributions from various processes. For 
reactions induced by the weakly bound projectile $^{6}$Li (Q = +1.47 MeV for 
the $\alpha$+$d$ breakup), it is natural to assume that an important 
contributor to the $\alpha$-particle cross sections is the breakup of $^{6}$Li
into $\alpha$ and $d$ fragments. Besides, other processes producing significant
$\alpha$-particle cross sections are also likely to occur. The processes that 
might contribute significantly to the inclusive $\alpha$-particle cross 
sections are:\\
i) Breakup of $^{6}$Li into  $\alpha$and $d$ fragments, which could be either 
direct or resonant (i.e. sequential) or both, where both fragments escape 
without being captured by the target, i.e., a no-capture breakup (NCBU) 
process,\\
ii) $\alpha$-particles resulting from $d$-capture by the target ($d$-ICF), 
following the breakup of $^{6}$Li into $\alpha$ and $d$, or a one-step $d$- 
transfer to the target,\\
iii) single-proton stripping from $^{6}$Li to produce unbound $^{5}$He that 
decays to an $\alpha$-particle plus a neutron,\\
iv) single-neutron stripping from $^{6}$Li to produce $\alpha$-unstable 
$^{5}$Li, that will subsequently decay to an $\alpha$-particle plus a proton, 
and\\
v) single-neutron pickup by $^{6}$Li to produce $^{7}$Li, which breaks into an 
$\alpha$-particleand a triton if $^{7}$Li is excited above its breakup 
threshold of 2.45 MeV.

 In order to understand the origin of the large inclusive $\alpha$-particle 
cross sections obtained in the reaction $^{6}$Li+$^{159}$Tb, measurements 
and/or theoretical calculations are necessary to estimate the 
contribution from each of the above processes.
\subsection{Breakup cross sections: CDCC calculations}
          To estimate the contribution from the NCBU process (i), exclusive 
measurements between the breakup fragments $\alpha$ and $d$ are needed. As only
inclusive measurements were taken in the present work, the NCBU cross 
sections have been estimated theoretically in the framework of the 
continuum-discretized-coupled channels (CDCC) method \cite{Ya86, Au87}. 
The CDCC calculations were performed with the coupled channels code FRESCO 
\cite{Th88} (version frxx.09j), by assuming $^{6}$Li to have an $\alpha$+$d$ 
cluster structure for its bound and continuum states. Following 
Ref.\cite{Ke98}, the $\alpha$-$d$ continuum was discretized into a series of 
equally spaced momentum bins, each of width $\Delta$k = 0.25 fm$^{-1}$ 
in the range 0.0 $\le$ k $\le$0.75 fm $^{-1}$, corresponding to the $^{6}$Li 
excitation energy of 1.47 $\le$E$_{x}$ $\le$10.27 MeV with respect to the 
$^{6}$Li ground state energy. The contribution from higher excited states is 
expected to be negligible. Each momentum bin was treated as an excited state of
$^{6}$Li nucleus with excitation energy equal to the mean energy of the bin and
having spin J and parity (-1)$^{L}$. The angular momenta are related by 
{\bf{J}}={\bf{L}}+{\bf{s}}, where {\bf{s}} is the spin of the $d$ and 
{\bf{L}} is the relative angular momentum of the $\alpha$$-$$d$ cluster system. 
In the calculations, L is limited to 0, 1, and 2. The contribution from higher 
L is negligible. Couplings to the 3$^{+}$ (E$^{*}$= 2.18 MeV), 2$^{+}$ 
(E$^{*}$=4.31 MeV) and 1$^{+}$ (E$^{*}$=5.65 MeV) resonant states as well as 
couplings to the non-resonant $\alpha$+$d$ continuum were included in the 
calculations. In order to avoid double counting, the bin width was suitably 
modified in the presence of resonant states. The $\alpha$+$d$ binding 
potentials were taken from Ref.\cite{Ku72}. The cluster-folding model potentials
for the interactions, $\alpha$$-$target and $d$$-$target were evaluated 
at 2/3 and 1/3 of the incident energy of the $^{6}$Li beam, respectively. As no 
experimental elastic scattering angular distribution data for 
$\alpha$+$^{159}$Tb and $d$+$^{159}$Tb reactions are available in the 
literature, the global optical model potential parameters \cite{Av94, An06} 
were used in describing the interactions at the corresponding energies. The 
couplings from the ground state to continuum and continuum to continuum states 
were included in the calculations. Both Coulomb and nuclear couplings were 
incorporated. The results of the NCBU cross sections thereby calculated are 
plotted in Fig.3 by the dash-dotted curve, and they are seen to largely 
underestimate the measured $\alpha$-particle cross sections. This shows that 
the $\alpha$-particles from sources other than breakup are important and need 
to be accounted for. This feature has also been observed for other heavy 
systems, such as $^{6,7}$Li+$^{208}$Pb \cite{Sig03} and $^{6}$Li+$^{209}$Bi 
\cite{Santra12}.
\subsection{Contribution of $\alpha$-particle cross sections from $d$-ICF 
process}
           The $\alpha$-particle cross sections resulting from the $d$-capture
by the $^{159}$Tb target ($d$-ICF, process (ii)), followed by $xn$ evaporation,
were determined from the $\gamma$-ray spectra recorded in the fusion cross 
sections measurement of the $^{6}$Li+$^{159}$Tb reaction \cite{Pra11}.  The 
cross sections of the resulting residual nuclei $^{160}$Dy, $^{159}$Dy, and 
$^{158}$Dy were already reported in Ref. \cite{Pra11}.  However, for the sake 
of convenience, the cross sections of the $\alpha xn$ channels, following the 
$d$-capture ICF, along with the total $d$-capture cross sections 
($\Sigma${$\alpha xn$}) are plotted in Fig.4. As already mentioned in the 
earlier work, the ICF cross sections thus measured also include contributions 
due to the $d$-transfer from $^{6}$Li to $^{159}$Tb, if any, since in the 
$\gamma$-ray measurement it was not possible to distinguish between the two 
types of events.

            Also, the single-proton stripping process (iii) 
$^{159}$Tb($^{6}$Li,$^{5}$He)$^{160}$Dy (Q = +2.836 MeV), if it occurs, will 
lead to the $^{160}$Dy nuclei in excited states. The $^{160}$Dy nuclei 
following the $1p$-stripping process will then decay by $xn$ evaporation to 
produce Dy-isotopes and will be included in the ${\alpha}xn$ channel cross 
sections of the $d$-ICF process. 

\begin{figure}
\vspace{-120pt}
\begin{center}
\includegraphics[scale=0.7]{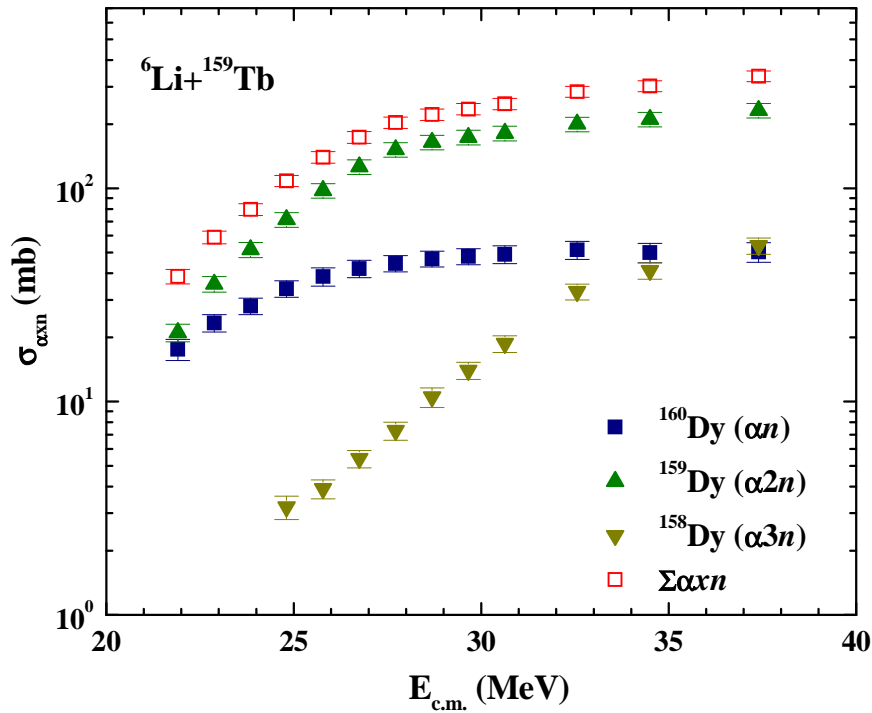}
\end{center}
\vspace{-120pt}
\caption{\label{fig:Fig4} (Color online)  
Cross sections of $^{160}$Dy, $^{159}$Dy and $^{158}$Dy nuclei produced in the
reaction $^{6}$Li+$^{159}$Tb. The hollow points are the sum of the cross
sections of the three Dy isotopes, i.e. the total $d$-ICF cross sections for the
reaction. See Sec.3.2 for details.} 
\end {figure}
\subsection{Contribution of $\alpha$-particle cross sections from $n$-transfer
processes} 
The contributions due to the processes (iv) and (v), i.e. the single-neutron 
stripping reaction $^{159}$Tb ($^{6}$Li,$^{5}$Li)$^{160}$Tb (Q = +0.711 MeV), 
and the single-neutron pickup reaction $^{159}$Tb($^{6}$Li,$^{7}$Li)$^{158}$Tb 
(Q = +0.883 MeV), leading to the $\gamma$-rays of $^{160}$Tb and $^{158}$Tb 
respectively, have been measured from the $\gamma$-ray spectra \cite{Pra11} of 
the $^{6}$Li+$^{159}$Tb reaction. For the former case, the production cross 
section of the 63.68 keV (1$^-$) state of $^{160}$Tb was obtained from the 
measured cross section of the 63.68 keV $\gamma$-ray, after correcting for its 
internal conversion coefficient ($\alpha_{T}$) of 15.1. In the $\gamma$-ray
spectra, this was the only $\gamma$-ray of $^{160}$Tb that could be identified.
Besides, because this is a fairly low energy $\gamma$-ray, special care was 
taken to estimate the area under this $\gamma$-ray peak. The 63.68 keV 
$\gamma$-ray is an E2 transition that feeds the ground state of $^{160}$Tb. As 
this $\gamma$-ray has a fairly large internal conversion coefficient, the 
reliability of the cross sections of $^{160}$Tb may be questioned. 
The large value of $\alpha_{T}$ = 15.1, 
though theoretically calculated, is expected to be a reliable estimate since 
theoretically calculated values of $\alpha_{T}$, especially for E2 transitions,
are known to agree well with the experimentally measured values. For example, 
the measured value of $\alpha_{T}$ for the 75.26 keV transition in $^{160}$Gd 
is 7.41$\pm$0.21 while  the calculated value varies between 7.24 and 7.51; the
measured $\alpha_{T}$ value for the 73.39 keV transition for $^{164}$Dy is 
8.92$\pm$0.19 while the calculated value varies between 8.80 and 9.12; the 
measured $\alpha_{T}$ value for the 53.2 keV transition for $^{230}$Th is 
229$\pm$7 and the calculated value lies between 227.6 and 234.2 \cite{Rama02}.  
Also, although the cross section of the 63.68 keV $\gamma$-ray is small, the 
corresponding peak in the $\gamma$-spectrum is fairly clean, thereby yielding 
$\gamma$-ray cross section with small uncertainty. Nevertheless, an uncertainty
of 10$\%$ in the theoretical value of $\alpha_{T}$ has been assumed while 
obtaining the cross sections of the $^{160}$Tb nuclei shown in Fig.5. It should
be emphasized here that in this method of extraction of $n$-stripping cross 
sections from the $\gamma$-ray spectra, the contribution of transfer to the 
ground state of $^{160}$Tb cannot be determined. So within the constraints of 
the present technique, only excited state transfer cross sections could be 
obtained.
\begin{figure}
\vspace{-100pt}
\begin{center}
\includegraphics[scale=0.6]{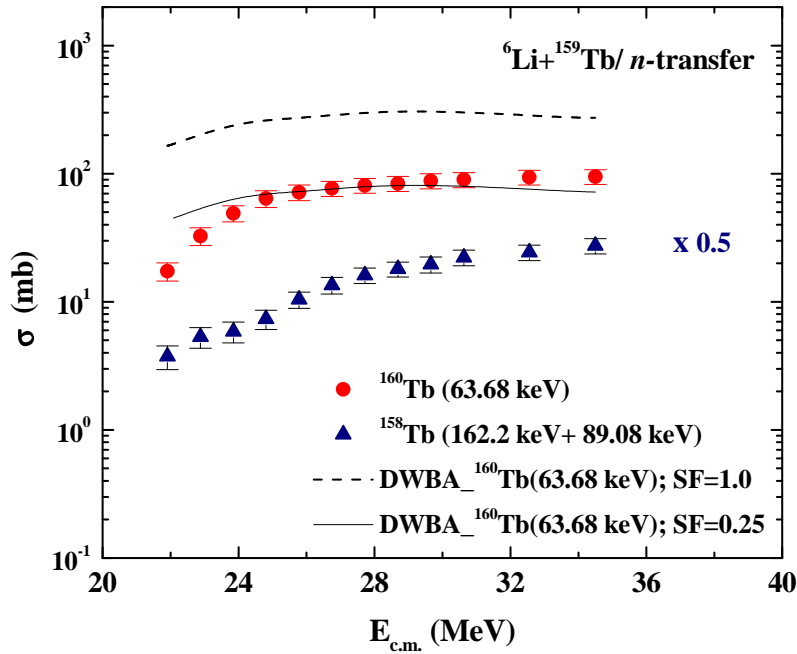}
\end{center}
\vspace{-100pt}
\caption{\label{fig:Fig5} (Color online)  
Measured cross sections of $1n$-stripping and $1n$-pickup processes populating
the excited states of $^{160}$Tb and $^{158}$Tb nuclei respectively, in the 
reaction $^{6}$Li+$^{159}$Tb. The dashed and solid curves are the DWBA 
calculated cross sections, with spectroscopic factors of 1 and 0.25, 
respectively, assumed for the 63.68 keV state. } 
\end{figure}

Similarly, the total cross sections of the $1n$-pickup process populating the 
excited states of $^{158}$Tb nuclei were obtained by summing the measured cross 
sections of the 162.2 keV and 89.08 keV $\gamma$-rays after appropriate 
correction for their respective internal conversion factors, and these are 
shown in Fig.5. In this case also, the ground state transfer cross sections 
could not be determined.

          In order to compare the measured cross sections for the $n$-stripping
process with theory, we attempted to calculate the cross sections 
for the single $n$-transfer to the first excited 63.68 keV (1$^{-}$) state of 
$^{160}$Tb nuclei for the five bombarding energies of 23, 25, 27, 30 and 35 
MeV. The transfer cross sections were calculated in the distorted wave Born 
approximation (DWBA) framework, using the computer code FRESCO. The 
$n$$-$$^{159}$Tb and $n$$-$$^{5}$Li binding potentials were taken from Refs. 
\cite{Be69} and \cite{Co67} respectively. The required potential parameters for 
the entrance channel $^{6}$Li+$^{159}$Tb, the exit channel $^{5}$Li+$^{160}$Tb 
and the $^{5}$Li+$^{159}$Tb core-core interaction were taken to be the global 
optical model potential parameters of Ref. \cite{Co82} with modifications such 
that these potentials fit the measured elastic scattering angular distributions 
at the five bombarding energies of 23, 25, 27, 30 and 35 MeV. Depth parameters 
have been adjusted to reproduce the binding energy of the neutron to the core 
$^{159}$Tb. The spectroscopic factor (SF) for $^{6}$Li$\rightarrow$$^{5}$Li 
+$n$ was taken from Ref. \cite{Co67}. The experimental SF for the 63.68 keV 
(1$^{-}$) state in $^{160}$Tb is not available in the literature. Nevertheless,
the transfer cross sections have been calculated by assuming the SF to be 1.0 
for the 63.68 keV (1$^{-}$) state. The cross sections thereby calculated are 
shown by the dashed curve in Fig.5, and they 
are seen to largely over-predict the measured cross sections. It was found that
the DWBA calculations done with a SF of 0.25 gave an overall fit to the 
measured cross sections at the higher energies. The resulting calculations are 
shown in the figure by the solid curve. 

         Due to the unavailability of relevant SFs, no better DWBA 
calculation could be done for the $1n$-stripping process. Therefore, no further
attempt was undertaken to calculate the cross sections for the $1n$-pickup 
process.
\subsection{Total contribution to measured $\alpha$-particle cross sections 
from various processes}
\begin{figure}
\vspace{-100pt}
\begin{center}
\includegraphics[scale=0.6]{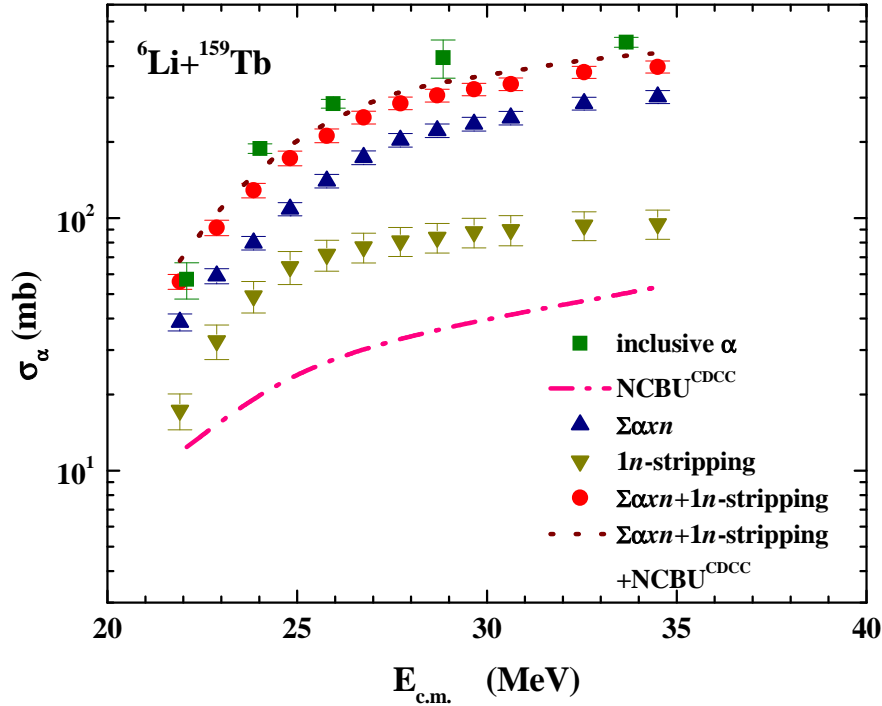}
\end{center}
\vspace{-80pt}
\caption{\label{fig:Fig6} (Color online)  
Contributions to the $\alpha$-particle cross sections originating from various 
processes in the reaction $^{6}$Li+$^{159}$Tb. The squares are the measured 
inclusive $\alpha$-particle cross sections. The up triangles are the 
$\Sigma$$\alpha xn$ channels, corresponding to the $d$-ICF process (including 
$d$-transfer and $1p$-stripping, if any). The down triangles are the 
contributions from the $1n$-stripping process (excluding ground state transfer),
corresponding to the instantaneous decay of the resulting $\alpha$-unstable 
$^{5}$Li nuclei into $\alpha $ and $p$. The circles are the sum of the $d$-ICF
and the $1n$-stripping cross sections. The cross sections resulting from the 
breakup of $^{6}$Li into $\alpha$ and $d$ (NCBU process), as determined from 
the CDCC calculations are shown by the dash-dotted line. The dotted curve shows
the trend of the cross sections, if the calculated NCBU cross sections are 
added to the measured cross sections of $d$-capture ICF and $1n$-stripping 
processes.}
\end{figure}

The measured $d$-ICF (i.e.$\Sigma \alpha xn$) cross sections, the $1n$-stripping
cross sections (excluding the ground state transfer contribution) and the sum of
the cross sections from these two processes are compared with the measured 
inclusive $\alpha$-particle cross sections in Fig.6. The calculated 
NCBU cross sections are also shown in the figure by the dash-dotted line. It is
observed that the $1n$-stripping cross sections (excluding the ground state
transfer contribution) are much larger than the calculated NCBU cross sections,
in contradiction to that reported for the $^{6}$Li+$^{209}$Bi reaction 
\cite{Santra12} but in agreement with the observation of Luong {\em{et al.}} 
\cite{Luong11, Luong13}. It had been mentioned in Ref. \cite{Santra12} 
that the measured exclusive cross sections of $\alpha$+$p$ in the 
$^{6}$Li+$^{209}$Bi reaction, following $n$-stripping of $^{6}$Li, are possibly 
the lower limit. This may be because the detector configuration used to measure
the $\alpha$+$p$ breakup cross section 
\cite{Santra09} did not cover the whole range of the relative momentum, thereby
leading to the underestimation of the cross section. Though the relative 
importance of reaction mechanisms largely depends on the target-projectile 
combination, the present observations, in conjunction with those reported in 
Refs. \cite{Luong11, Luong13}, in fact do show that the $n$-stripping process 
is more important than the NCBU process in $^{6}$Li-induced reactions with 
targets such as $^{159}$Tb, $^{207,208}$Pb and $^{209}$Bi. 

It can be seen from Fig.6 that the sum of the cross sections resulting from 
$d$-ICF (including $d$-transfer and $1p$-stripping, in any) and $1n$-stripping 
(excluding ground state transfer) reactions, shown by the solid circles, lie 
very close to the measured total inclusive $\alpha$-particle cross sections. 
Here the yield of the $\alpha$-particles due to the ground state transfer in 
the $1n$-stripping process, and also following the breakup of $^{7}$Li nuclei 
produced via the $n$-pickup process have not been considered.
The $\alpha$-breakup threshold of $^{7}$Li is 2.45 MeV, and hence $^{7}$Li 
nuclei can breakup only if they are excited above 2.45 MeV. Therefore, the 
$1n$-pickup reaction will contribute to the total $\alpha$-particle cross 
sections, depending on the excitation energy of the $^{7}$Li nuclei. At lower 
bombarding energies, this process may not be a significant contributor. But at 
higher bombarding energies, the $^{7}$Li nuclei may be excited to energies 
above the breakup threshold, thereby resulting in a small contribution. However,
it is obvious from the figure that for this reaction, over the energy range of 
the present measurement, the $1n$-pickup process is certainly not a very 
significant contributor to the total $\alpha$-particle cross sections. The 
dotted curve in the figure shows that if we add the CDCC calculated NCBU cross 
sections to the measured $d$-ICF and $1n$-stripping cross sections, the 
inclusive $\alpha$-particle cross sections are nearly reproduced. Thus, 
the $d$-ICF (including $d$-transfer and $1p$-stripping, if any) and the 
$1n$-stripping processes are the dominant contributors, with the NCBU process 
being a relatively small contributor, to the total $\alpha$-particle cross 
sections in the $^{6}$Li+$^{159}$Tb reaction at energies around the Coulomb 
barrier. 
\section{Summary}
    In summary, the inclusive $\alpha$-particle cross sections for the reaction 
$^{6}$Li+$^{159}$Tb have been measured at energies around the Coulomb barrier.
The NCBU cross sections calculated using the CDCC formalism are found to be 
only a small fraction of the inclusive $\alpha$-particle cross sections. Other 
reaction mechanisms contributing to the large $\alpha$-particle cross sections 
have been disentangled, using data from our earlier work \cite{Pra11} based on 
an entirely different technique, {\em{e.g.}} the $\gamma$-ray method. The $1n$- 
stripping cross sections are found to be much larger than the calculated cross 
sections of the NCBU process, in contradiction to the observation reported for 
the reaction $^{6}$Li+$^{209}$Bi \cite{Santra12}. The $d$-ICF, including 
$d$-transfer and $p$-stripping if any, and the $1n$-stripping processes are 
found to be the dominant contributors to the total $\alpha$-particle cross 
sections in the $^{6}$Li+$^{159}$Tb reaction. However, due to the lack of 
appropriate spectroscopic factors, proper DWBA calculations could not be 
performed. Experiments aimed at measuring such spectroscopic factors need to be 
carried out in the near future. Besides, as transfer induced breakup seems to 
be an important process in reactions with loosely bound projectiles, both 
inclusive and exclusive measurements in other systems, especially lighter 
systems and systems involving halo and skin nuclei, would be very valuable to 
obtain a clear picture of the transfer-breakup process. Identification and 
subsequent determination of the absolute cross sections of different multi-step
reaction processes involved in reactions with weakly bound nuclei may pave the 
way for the theorists to come up with a proper theoretical description of such 
processes, which is indeed a challenging task. 

\begin{acknowledgments}
We are grateful to Prof. B.K. Dasmahapatra for valuable discussions and advices
at various stages of the work. We thank Mr. P.K. Das for his earnest technical 
help during the experiment. We would also like to thank the accelerator staff 
at the BARC-TIFR Pelletron Facility, Mumbai, for their untiring efforts in
delivering the beams.
\end{acknowledgments}

\end{document}